\documentclass[aps,prl,twocolumn,superscriptaddress]{revtex4}

\usepackage{mathrsfs,amssymb,graphics,subfigure,threeparttable}
\usepackage{grap hicx}
\usepackage{amssymb}
\usepackage{enumerate}
\usepackage{amsmath}
\usepackage{fancyhdr}
\usepackage{bm}
\usepackage[squaren]{SIunits}

\begin{document}


\title{High quality electron bunch generation using a longitudinal density-tailored plasma-based accelerator in the three-dimensional blowout regime}


\author{X. L. Xu}
\email[]{xinluxu11@ucla.edu}
\affiliation{University of California, Los Angeles, California 90095, USA}
\author{F. Li}
\affiliation{Department of Engineering Physics, Tsinghua University, Beijing 100084, China}
\author{W. An}
\affiliation{University of California, Los Angeles, California 90095, USA}
\author{P. Yu}
\affiliation{University of California, Los Angeles, California 90095, USA}
\author{W. Lu}
\email[]{weilu@mail.tsinghua.edu.cn}
\affiliation{Department of Engineering Physics, Tsinghua University, Beijing 100084, China}
\affiliation{IFSA Collaborative Innovation Center, Shanghai Jiao Tong University, Shanghai 200240, China}
\author{C. Joshi}
\affiliation{University of California, Los Angeles, California 90095, USA}
\author{W. B. Mori}
\affiliation{University of California, Los Angeles, California 90095, USA}


\date{\today}

\begin{abstract}
The generation of very high quality electron bunches (high brightness and low energy spread) from a plasma-based accelerator in the three-dimensional blowout regime using self-injection in tailored plasma density profiles is analyzed theoretically and with particle-in-cell simulations. The underlying physical mechanism that leads to the generation of high quality electrons is uncovered by tracking the trajectories of the electrons as they cross the sheath and are trapped by the wake. Details on how the intensity of the driver and the density scale-length of the plasma control the ultimate beam quality are described. Three-dimensional particle-in-cell simulations indicate that this concept has the potential to produce beams with peak brightnesses between $10^{20}$ and $10^{21}~\ampere/\meter^{2}/\rad^{2}$ and with absolute projected energy spreads of $\sim 0.3~\mathrm{MeV}$ using existing lasers or electron beams to drive nonlinear wakefields.   
\end{abstract}

\pacs{}

\maketitle


Research in Plasma-based acceleration (PBA) driven by a laser pulse or a relativistic electron beam is very active \cite{joshi2003plasma} because the large accelerating gradients in plasma wave wakefields may lead to ultra compact accelerators. PBA is also capable of self-generating electron bunches that contain a significant amount of charge (Q), have short durations ($\tau$) and low normalized emittance ($\epsilon_n$). These beam quantities are often combined into the normalized beam brightness $B_n=2I/\epsilon_{n}^2$ where $I=Q/\tau$ is the current. Ultra-high beam brightnesses are needed in accelerator-based x-ray light sources \cite{barletta2010free}. While PBA experiments have produced useful beams, they have not produced beams with the necessary brightnesses and energy spreads needed to drive an X-ray free-electron-laser (X-FEL) or the charge and emittance needed as an injector for a future linear collider. 

The electron bunches needed to load plasma wave wakefields are very short and they also need to be synchronized with the driver. Therefore, self-injection has been actively investigated. Self-injection of electrons and its threshold into nonlinear plasma waves in uniform plasmas has been studied in simulations \cite{PhysRevLett.93.185002, gordienko2005scalings, PhysRevSTAB.10.061301, kalmykov2009}. This process does not appear to be capable of generating the high quality beams needed for coherent radiation sources or a linear collider \cite{mangles2004monoenergetic, geddes2004high, faure2004laser}. Therefore there has been much recent theoretical and computational work on methods for generating  high brightness beams through controlled injection into plasma wave wakes \cite{chen2006electron, PhysRevLett.108.035001, PhysRevLett.111.015003, PhysRevLett.112.035003, bulanov1998pre, suk2001, PhysRevLett.93.185002, kalmykov2009, kostyukov2009, PhysRevLett.111.085005}. These ideas fall into three categories. In one, electrons are born inside the wake through field ionization where the wake potential is near a maximum that eases the trapping threshold \cite{PhysRevLett.98.084801, PhysRevLett.104.025003}. There are now numerous variations of this idea in which the injection and wake excitation are separated \cite{PhysRevLett.108.035001, PhysRevLett.111.015003, PhysRevLett.112.035003}. In the second, the effective phase velocity of the wake is slowed down either by a density transition from high to low density \cite{bulanov1998pre, suk2001}, or through an expanding bubble from the evolution of a laser driver \cite{PhysRevLett.93.185002, kalmykov2009, kostyukov2009, PhysRevLett.111.085005}. In the third, one or more laser pulses are used to trigger injection inside one plasma wake bucket \cite{PhysRevLett.76.2073, PhysRevLett.79.2682, PhysRevLett.102.065001}. Simulations have shown that these ideas provide beams with a variety of different Q, $\tau$, $\epsilon_n$, and energy spread, $\sigma_{\gamma}$.

\begin{figure}[bp]
\includegraphics[width=0.5\textwidth]{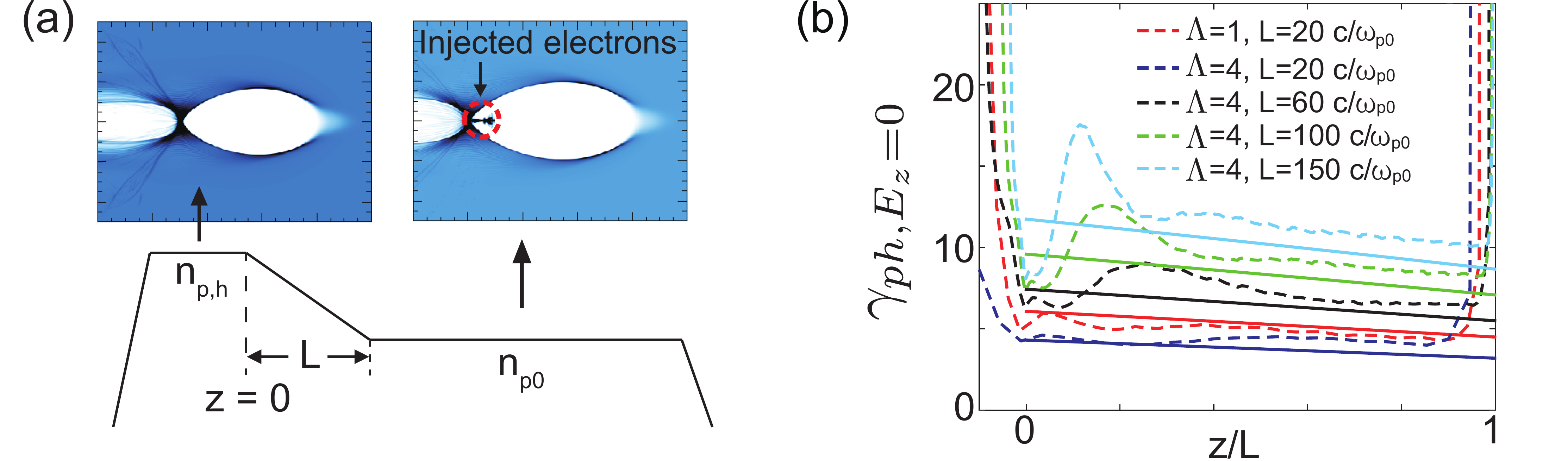}
\caption{\label{fig: } (a) Schematic of density downramp injection. The plasma density decreases linearly from $n_{p,h}$ at $z=0$ to $n_{p0}$ at $z=L$. (b) Evolution of the phase velocity $\gamma_{ph,E_z=0}$ from Eq. (1) (solid lines) and  3D PIC simulations (dashed lines). The parameters are: $n_{p,h}=1.5 n_{p0}, \gamma_b=2500, n_b=16n_{p0}, \sigma_z=0.7\frac{c}{\omega_{p0}}$, $\sigma_r=0.25\frac{c}{\omega_{p0}}$ when $\Lambda=1$ and $\sigma_{r}=0.5 \frac{c}{\omega_{p0}}$ when $\Lambda=4$. A longer simulation box is used when $\Lambda=4$. }
\end{figure}

As first pointed out by Katsouleas \cite{PhysRevA.33.2056}, the phase velocity  of a wake driven by a particle beam moving with a constant velocity $v_d$ in a density gradient will change due to the density dependence of the wavelength. The phase of the plasma wave wake behind the driver can be written as $\phi(z,t) = \omega_p(z)\left (z/v_d - t \right)$, where $\omega_p(z)$ is the local plasma frequency. Thus the phase velocity of the wake is
\begin{align}
v_\phi (z, t ) = \frac{v_d}{ 1-\left( \mathrm{d}\omega_p/\mathrm{d}z \right)\omega_p^{-1} (v_d t-z)}
\end{align}

This can be used to increase the phase velocity (upramp) or decrease the phase velocity (downramp). The concept of using variations of the plasma density to trigger injection was subsequently proposed in gradual \cite{bulanov1998pre} and sudden \cite{suk2001} density transitions from a high density plasma to a low density plasma. These analyses were based on one-dimensional (1D) arguments. There have also been some recent results from multi-dimensional simulations \cite{suk2001, geddes2008densityramp, gonsalves2011tunable, buck2013density-ramp, grebenyuk2014beam}. However no analysis of the phase space dynamics of the injected electrons needed to understand why and how ultra bright electron beams can be generated was provided. 
 
In this Letter, we analyze the 3D self-injection in density downramps from wakes excited in the nonlinear blowout regime using theory and OSIRIS \cite{fonseca2002high} simulations. We find that beams with unprecedented brightnesses ($\gtrsim 10^{20}~\ampere/\meter^{2}/\rad^{2}$) can be generated under the appropriate conditions. The phase velocity is controlled by varying the blowout radius using a density ramp and emittance is controlled because of the defocusing fields on the electrons as they converge back to the axis in the density spike in the rear of the bubble. 

The processes behind the injection and generation of ultra bright electron beams are clearly illuminated by tracking particles of interest. For clarity in interpretation of the physics we use a non-evolving ultra-relativistic electron beam to produce the wake; however, when evolving beams or lasers are used similar results are obtained. By adjusting the magnitude of the plasma density gradient and the driver intensity, one can control the expansion rate of the blowout radius so that electron trapping occurs. We show that downramp injection in the blowout regime can generate the  brightnesses and energy spreads needed to drive an XFEL into saturation with a much shorter undulator at nm or smaller wavelengths; and that this scheme can generate 100's of pC of charge with normalized emittances less than 50nm making it a possible injector for a future linear collider. None of the ionization based injection schemes nor previous work on downramp injections has indicated that such beam parameters can be produced.  

The basic idea is illustrated in Fig. 1(a). In the OSIRIS simulations used for Figs. 1-4,  we use $512 \times 512\times 320$ cells in the $x, y$, and $z$ directions,   cell sizes of $\frac{1}{32}\frac{c}{\omega_{p0}}$ in each direction, and 4-8 particles per cell is used for the plasma electrons. When a high current electron bunch propagates through plasma, a highly nonlinear plasma wave structure can be excited if the bunch peak density $n_b$ exceeds the plasma density $n_p$ \cite{PhysRevA.44.R6189, PhysRevLett.96.165002, lu2006nonlinearPoP} and the peak normalized charge per unit length, $\Lambda \equiv 4\pi r_e\int_0^{r\gg \sigma_r} rdr n_b$ exceeds unity, where $\sigma_r$ is the spot size of the beam and $r_e$ is the classical electron radius. For $\Lambda \gg 1$,  the Coulomb force of the drive electron bunch ``blows out" the plasma electrons which then form a thin sheath surrounding a bubble-like region that contains only the ``immobile" ions. In the laser driver case,  a similar bubble structure is formed if the normalized vector potential $a_0 \equiv \frac{eA_0}{mc^2} \gg 1$ where $A_0$ is the peak vector potential of the laser \cite{mori1991blowout, pukhov2002laser, PhysRevLett.96.165002, lu2006nonlinearPoP}. 

\begin{figure}[bp]
\includegraphics[width=0.5\textwidth]{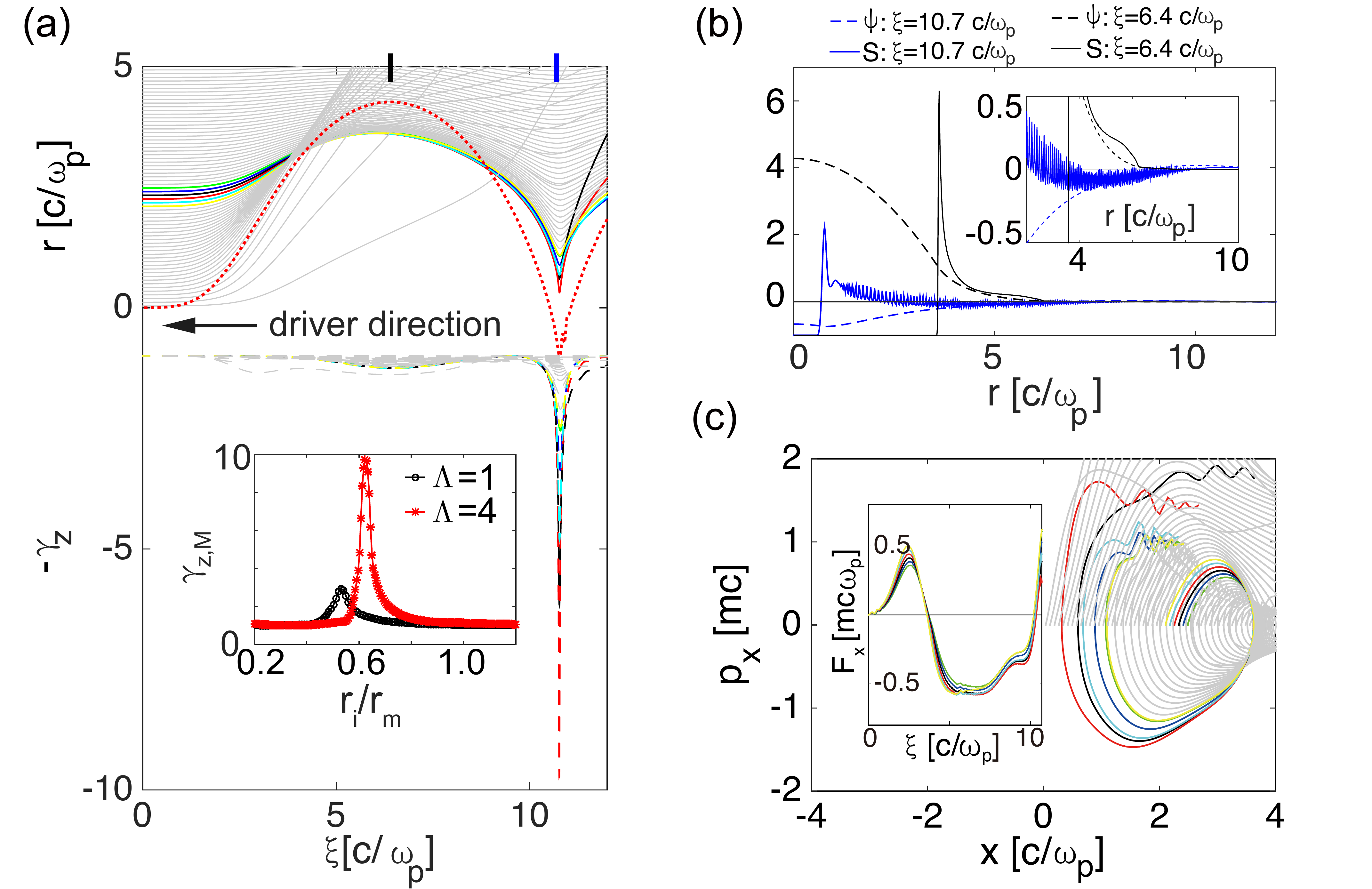}
\caption{\label{fig: } (a) The trajectories (solid lines) and the longitudinal velocities (dashed lines) of the electrons in a uniform plasma with $n_p=1.5 n_{p0}$. The inset shows the dependence of the maximum $\gamma_z$ of the electrons on the initial radius under different drivers. The red dotted line is the on-axis $\psi_0$.  (b) The source term $S$ and $\psi$ at different $\xi$. The inset enlarges the vertical scale to [-0.5, 0.5]. (c) The trajectories of the electrons in the transverse phase space. The inset shows the transverse forces experienced by the sheath electrons. The beam parameters are the same as the $\Lambda=4$ case in Fig. 1. The electrons are selected to have negligible motion in the $y$-direction.}
\end{figure}

In the blowout regime, the edge of the ion column is called the blowout radius, $r_b(\xi)$ (the radius is in cylindrical coordinates) where $\xi \equiv v_d t-z \approx ct-z$. The maximum value of $r_b$ is defined as $r_m$ and for a particle beam driver is given by $2\sqrt \Lambda c/\omega_p$ \cite{PhysRevLett.96.165002, lu2006nonlinearPoP}. When $r_m\gg c/\omega_p$ then $r_b(\xi)$ nearly maps out the ion column that resembles a nearly spherical bubble and the wavelength of the wake is therefore $\lambda_{wake} \approx 2 r_m \approx 4\sqrt\Lambda c/\omega_p$. The corresponding frequency is therefore $\omega_{NL}=\frac{\pi \omega_p}{2 \sqrt \Lambda}$ where we note that $\Lambda$ is independent of density. Therefore  when $\omega_p$ is replaced by $\omega_{NL}$ in the expression for the phase velocity the result is unchanged. For the velocity of the first density spike, we can replace $\lambda_{wake}$ for $v_d t-z$ in Eq. (1) to get $v_{\phi} \approx v_d \left(1-4\sqrt \Lambda \frac{\mathrm{d} \omega_p^{-1}}{\mathrm{d} z} \right)$. Below we track where $E_z=0$ and assume it behaves similar as the density spike where $E_z$ is a minimum. Within the density downramp $\frac{\mathrm{d} \omega_p^{-1}}{\mathrm{d} z} > 0$, so $v_{\phi}$ (and therefore $\gamma_\phi$) can be much reduced from $v_d$ as shown in Fig. 1(b) where $\gamma_{\phi,E_z=0}$ is plotted vs. the location in the density downramp for several values of $L$ and $\Lambda$.

In a region of gradual density decrease, i.e., $l\equiv \frac{n_p}{\mathrm{d}n_p/\mathrm{d}z}\gg \frac{c}{\omega_p}$, the motion of an electron before injection is similar to the motion in a uniform plasma. As pointed out in Refs. \cite{PhysRevLett.96.165002, lu2006nonlinearPoP}, in the blowout regime the trajectories of the plasma electrons vary significantly depending on their impact parameter $r_i$, i.e., the initial radius. As seen in Fig. 2(a), the electrons with small impact parameter $r_i \ll r_m$ are deflected by the driver whereas the electrons with large impact parameter $r_i \gg r_m$ are hardly perturbed. Only the electrons with $r_i \approx \kappa r_m$ form the high density narrow sheath of the wake, where $\kappa \approx \frac{1}{2}$ and its precise value depends on the intensity and profile of the driver and can be deduced from simulations \cite{PhysRevLett.96.165002, lu2006nonlinearPoP}. These electrons obtain large longitudinal forward velocity $\gamma_z \equiv (1-\beta_z^2)^{-\frac{1}{2}}$ when they reach the rear of the wake as shown by the dashed lines in Fig. 2(a). The dependence of the maximum longitudinal velocity of the electrons $\gamma_{z,M}$ on the impact parameter for different driver intensities are shown in the inset in Fig. 2(a). One can see a stronger driver tends to generate electrons with larger $\gamma_{z,M}$. When the driver propagates through a gradual density downramp, some electrons in the sheath can satisfy $\gamma_z \geq \gamma_{ph}$  as it is pulled back to the axis at which time it becomes injected. This selection mechanism determines the beam quality generated in density downramp injection. 

The longitudinal velocity of the electron can be described as  $\beta_z=1- \frac{2(1+\psi)^2}{1+(p_\perp/mc)^2 + (1+\psi)^2}$ \cite{mora1997kinetic}, where $\psi\equiv \frac{e}{mc^2} \left( \phi - A_z\right)$ is the wake potential and $\phi$ and $A_z$ are the scalar potential and the axial component of the vector potential, respectively. The wake potential $\psi$ obeys the Possion like equation $\nabla_\perp^2\psi=S\equiv-\frac{1}{n_p e}\left(\rho-\frac{J_z}{c}\right)$  \cite{mora1997kinetic}, where $S=-1$ inside the ion column. If $S\geq 0$ outside the ion column, then by integrating the Possion like equation it is straightforward to show that $\psi \geq 0$ at each transverse position, such as in the case for the black lines in Fig. 2(b). However at the very rear of the wake, $S<0$ for some $r$ outside the ion column which may lead to a negative $\psi$ inside the ion column, such as in the case for the blue lines in Fig. 2(b). At the very rear of the wake, $\psi$ can be very close to $-1$, and this is why the electrons always obtain the largest forward velocity at the very rear of the wake. 

The transverse force, $-e(\vec E + \frac {\vec v}{c} \times \vec B)_{\perp}$,  on a plasma electron at $r=r_b(\xi)$ can be written as $F_r = F_{d}  + F_{i}  + F_{e}$ \cite{PhysRevLett.96.165002, lu2006nonlinearPoP}, where $F_{d}=mc\omega_p(1-\beta_z)\frac{\Lambda(\xi)}{k_pr}$ is from the beam driver, $F_{i}=- mc\omega_p\frac{k_pr}{2}$ is from the ion cavity and $F_e= - mc\omega_p \frac{k_pr}{2}(1-\beta_z)\frac{\mathrm{d}^2\psi_0}{\mathrm{d}(k_p\xi)^2}$ is from the plasma electrons. Initially the force from the driver $F_{d}$ dominates this is what expels the plasma electrons outward forming the narrow sheath.  As the driver passes by the electrons, the focusing force is solely due to  $F_i+F_e$. For most of the wake, $\frac{\mathrm{d}^2\psi_0}{\mathrm{d}(k_p\xi)^2}\approx -\frac{1}{2}$ \cite{PhysRevLett.96.165002, lu2006nonlinearPoP}, thus $\frac{F_i+F_e}{r} <0$ and the plasma electrons are pulled back to the ions. However, at the very rear of the wake, the $|\frac{\mathrm{d}^2\psi_0}{\mathrm{d}(k_p\xi)^2}|$ term can be very large leading to $\frac{F_r}{r} >0$ [see the inset in Fig. 2(c)] and these electrons are decelerated in the transverse direction as they rush back to the axis, i.e., the amplitude of the radius and transverse momentum decrease simultaneously [see Fig. 2(c)]. These electrons have $\beta_z \sim 1$ so they can remain in phase with these defocusing fields therefore the impulse from $F_r$ can reduce the transverse momentum to a vey low value as shown in Fig. 2(c). This leads to a much lower emittance for the trapped electrons.

\begin{figure}[bp]
\includegraphics[width=0.5\textwidth]{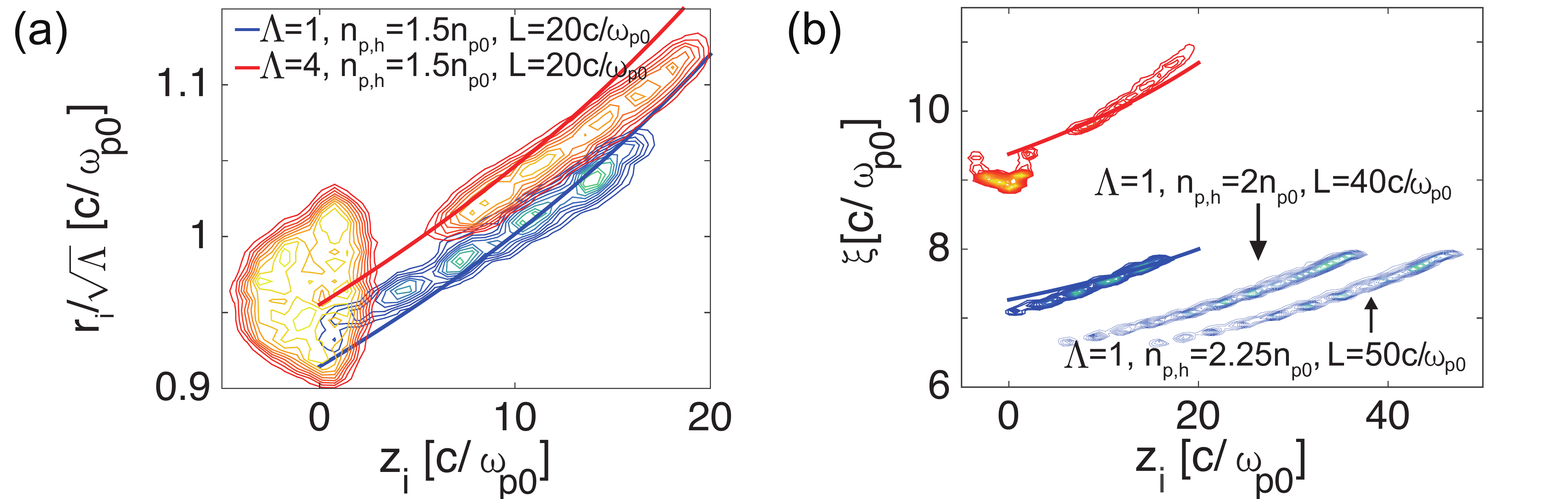}
\caption{\label{fig: } The normalized charge density distribution of the injected electrons in $r_i - z_i$ plane (a) and $\xi - z_i$ plane (b). The lines are from Eq. (\ref{eq: ri_zi}). The beam parameters are the same as in Fig. 1. }
\end{figure}

As shown above, the electrons in the sheath can obtain a large enough forward velocity to begin to move synchronously with the wake if $v_\phi$ is reduced enough in the density downramp region so that $v_z$ exceeds it. For a monotonically decreasing density,  after injection the electrons then gradually move forward relative to the rear of the wake due to the further expansion of the wake, i.e., they do not phase slip with respect to the driver but they do with the rear of the bubble. Meanwhile the impact parameter of the injected electrons increases because it depends on density. There is also a one-to-one mapping between $\xi$ (the axial location of the electron after the downramp region) and the initial longitudinal position $z_i$, which means the longitudinal phase mixing \cite{PhysRevLett.112.035003} is ideally suppressed, i.e., the electrons in a final longitudinal slice originate from the same initial longitudinal position, therefore an injected beam with low slice energy spread can be expected. The dependence of $r_i$ and $\xi$ on $z_i$ are therefore
\begin{align}
\frac{\mathrm{d}r_i}{\mathrm{d}z_i} \approx  \kappa\frac{\mathrm{d}r_m}{\mathrm{d}z_i}, \frac{\mathrm{d}\xi}{\mathrm{d}z_i} \approx \frac{\mathrm{d} \lambda_{wake}}{\mathrm{d}z_i}\approx 2\frac{\mathrm{d}r_m}{\mathrm{d}z_i}
\label{eq: ri_zi}
\end{align}
In Fig. 3, the dependence of $r_i$ and $\xi$ on $z_i$ for the injected electrons with different driver intensities are shown. Good agreement between OSIRIS PIC simulation results and Eq. (2) is found. When $\Lambda=1$, the electrons are injected continuously as the driver propagates in the ramp. When $\Lambda=4$, there is significant injection around the starting point of the ramp ($z_i=0$). In this first injection, the electrons from different $z_i$ mix together, leading to a large slice energy spread and large current. These injected electrons load the wake \cite{PhysRevLett.101.145002}, decreasing the longitudinal velocity of the subsequent electrons in the sheath and the injection ceases. As the beam propagates further into the low density region, $v_{\phi}$ decreases further and the beam loading effect weakens due to the injected electrons moving forward relative to the rear of the wake \cite{PhysRevLett.101.145002}, leading to a second injection phase. In this second injection phase, the electrons are then injected continuously as is the case when $\Lambda=1$. In Fig. 3(b), we show two more cases for $\Lambda=1$ with the same density gradient but different ramp lengths. Because $v_{\phi}$ of the wake is always higher at the beginning of the ramp for a linear profile [see Fig. 1(b)], the injection process is the same for the profiles with the same density gradient and minimum density but longer lengths. 

The current of the injected beam can be approximated as $I \approx (ec)2\pi r_i \Delta r_i n_p \frac{\mathrm{d}z_i}{\mathrm{d}\xi} \approx - (\pi ec)\Delta r_i n_p l$, where $2\pi r_i \Delta r_i n_p \mathrm{d}z_i$ is the injected particle number at $z_i$, $\mathrm{d}\xi$ is the length of the these electrons after injection, and we use relationship Eq. (2). The current can therefore be controlled by adjusting the driver intensity and the gradient of the downramp. An intense driver increases $\Delta r_i$. A steeper ramp (i.e., smaller $l$) makes injection easier by decreasing the $\gamma_z$ needed for injection which then increases $\Delta r_i$,  but it also simultaneously decreases $n_p l$, so it is necessary to use simulations to quantify the dependence of the current on the density scale length.

\begin{figure}[bp]
\includegraphics[width=0.5\textwidth]{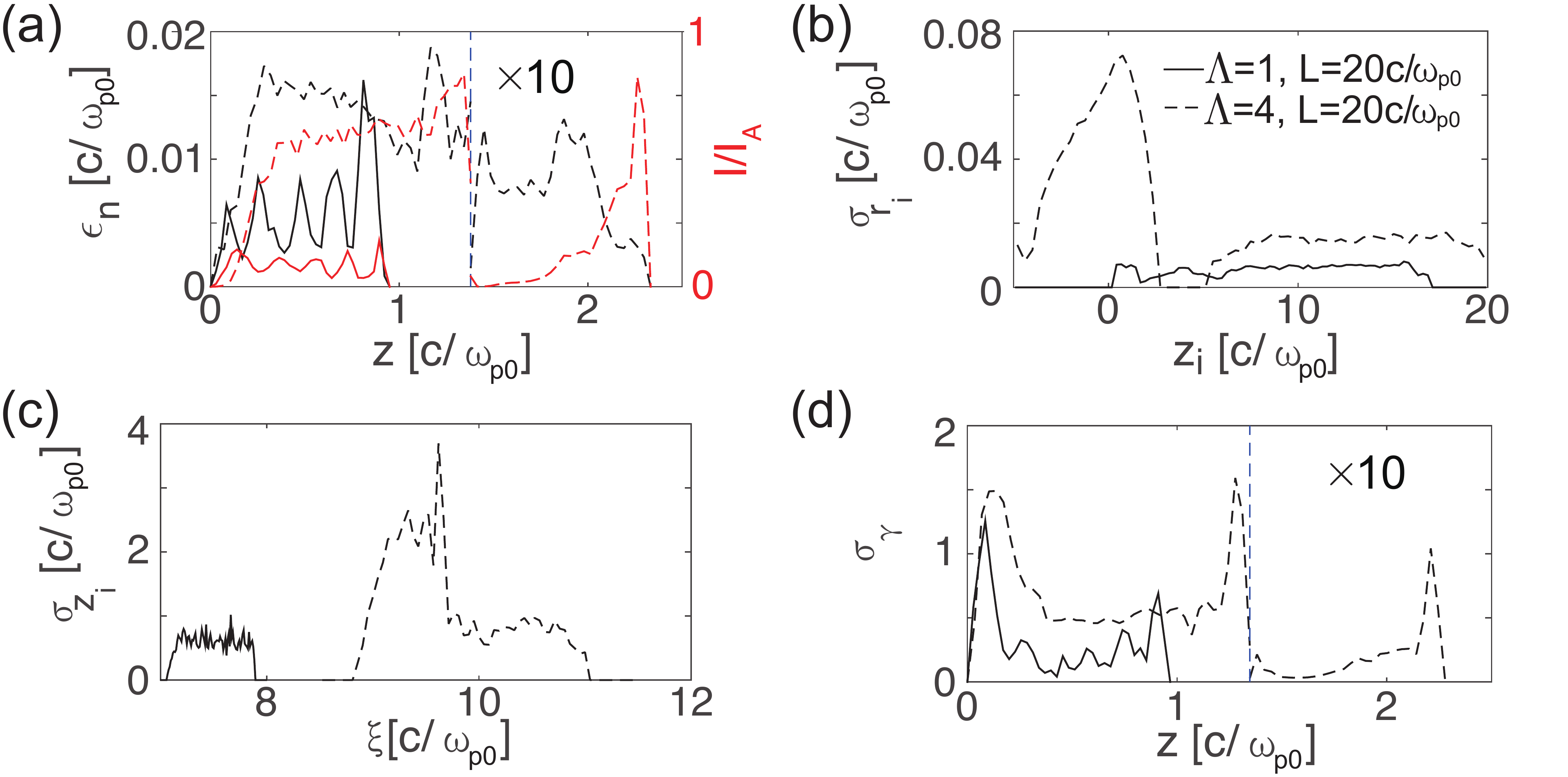}
\caption{\label{fig: } (a) The emittance and the current of the beam. (b) The dependence of $\sigma_{r_i}$ on $z_i$ and (c) the dependence of $\sigma_{z_i}$ on $\xi$ of the injected beam. (d) The slice energy spread of the beam. The beam is divided into 32 (64) slices when $\Lambda=1 (4)$. The mean energy is around 55 MeV.}
\end{figure} 

The slice emittance of the injected beam is determined by the value of the impact parameter $\Delta r_i$. Large $\Delta r_i$ tends to generate a beam with a larger emittance. So we can fine tune the parameters of the density profile and the driver to reduce the width of the impact parameter for injected electrons to generate a beam with a low emittance. The current, slice emittance and energy spread of the injected beam when $\Lambda=1$ and $\Lambda=4$ are shown in Fig. 4. In Fig 4(a) we can see that the emittance and the current are smaller when $\Lambda=1$ as compared to when $\Lambda=4$, this is consistent with the distribution of $\sigma_{r_i}$ (hereafter, $\sigma_{r_i}$ refers to the rms value of the subscript quantify for the injected particles) as shown in Fig. 4(b). In Fig. 4(c), the dependence of $\sigma_{z_i}$ on $\xi$ is shown, which shows that longitudinal mixing is ideally suppressed in the injected beam when $\Lambda=1$ and the second injected beam when $\Lambda=4$ corresponding to $\sigma_{z_i} \lesssim1$. But in the first injection for $\Lambda=4$ the electrons from different $z_i$ mix together where $\sigma_{z_i}\sim 3$. As a result, slice energy spreads as low as $\sigma_\gamma\lesssim0.5$ for the beam when $\Lambda=1$ and the second injected beam when $\Lambda=4$ are obtained; the slice energy spread for the first injected beam when $\Lambda=4$ is as high as $\sigma_\gamma\sim 3$. Note that in the simulations, in order to eliminate the numerical Cherenkov instabilities \cite{godfrey2013numerical, xu2013numerical} induced by the high current, relativistic drifting injected beam, the PIC code OSIRIS with FFT/Finite-difference solver \cite{Yu2015} is used.


The effect of plasma temperature and asymmetric driver on the injected beam quality are also studied with simulations. For the $\Lambda=4, L=20 c/\omega_{p0}$ case, the emittance is larger by a factor 1.3 for a 40 eV plasma temperature as compared with a cold plasma. However, the current and the energy spread vary little. We also find that the aspect ratio of the driver should be kept below 1.1 to keep the beam quality unchanged because the deceleration of the sheath electrons in the transverse directions is related to the radial symmetry geometry.  

Each PIC simulation corresponds to an infinite set of physical parameters with the same normalized parameters and beam shapes but different absolute plasma density. The brightnesses of both the drive and output beams are proportional to the corresponding plasma density. For example, for the $\Lambda=1$ and $4$ cases presented, the peak brightness of the output beams correspond to $30$ and $15(n_{p0}[\centi\meter^{-3}]) \ampere/\meter^{2}/\rad^{2}$ respectively. Current state-of-the-art electron beams can operate in the blowout regime in densities as high as $10^{20}~\centi\meter^{-3}$ \cite{Brendan}, indicating brightnesses greater than $10^{21}~\ampere/\meter^{2}/\rad^{2}$ are possible. A laser driver can also be used. Here we present results from an OSIRIS simulation of an 800 $\nano\meter$ circularly-polarized laser pulse with $a_0=2\sqrt{2}, w_0=5.5~\micro\meter$ and $\tau_{FWHM}=25~\femto\second$ exciting a wake in a plasma with $n_{p0}=10^{19}~\centi\meter^{-3}$. The laser is focused at $z_{plasma}=-0.025\milli\meter$, where $z_{plasma}=0~\milli\meter$ is the start of the downramp. The plasma density is decreased from $1.5n_{p0}$ to $n_{p0}$ in $28~\micro\meter$. The simulation window has a dimension of $40.6 \times 40.6 \times 35.6~\micro\meter$ with $1600\times 1600\times 1400$ cells in the $x, y$ and $z$ directions, respectively. In this example an electron beam with a peak brightness of $1.8\times10^{20}\ampere/\meter^{2}/\rad^{2}$ (8 $\mathrm{kA}$ current and 9 $\mathrm{nm}$ emittance) and $\lesssim 0.15$ MeV slice energy spread is generated. A positive energy chirp long $z$-direction is presented immediately after the ramp due to the mapping between $z_i$ and $\xi$ [see Fig. 3(b)]. At some specific acceleration distance, this positive chirp is removed due to the negative chirp of the acceleration gradient and the projected energy spread of the beam is very low. At $z_{plasma}=0.18~\milli\meter$, the projected rms energy spread of the injected beam between $z=1.2~\micro\meter$ and $z=3.0~\micro\meter$ is only 0.27 MeV with a mean energy 56.5 MeV. 
\begin{acknowledgments}
This work was supported by the National Basic Research Program of China Grant No. 2013CBA01501, NSFC Grants No. 11425521, No. 11535006, No. 11375006, and No. 11475101, Thousand Young Talents Program, DOE Grants No. DE-SC0010064, No. DE-SC0014260, and NSF Grants No 1500630, No. ACI-1339893, No. ACI-1440071, No. PHY-1415386. The simulations were performed on the UCLA Hoffman 2 and Dawson 2 Clusters, and the resources of the National Energy Research Scientific Computing Center and the Blue Waters. 

X. L. Xu and F. Li contributed equally to this work.
\end{acknowledgments}

\bibliography{refs_xinlu}

\end{document}